\documentclass[12pt,onecolumn]{article}

\usepackage{PRIMEarxiv}

\usepackage[utf8]{inputenc} 
\usepackage[T1]{fontenc}    
\usepackage{hyperref}       
\usepackage{url}            
\usepackage{booktabs}       
\usepackage{amsfonts}       
\usepackage{nicefrac}       
\usepackage{microtype}      
\usepackage{lipsum}
\usepackage{fancyhdr}       
\usepackage{graphicx}       
\graphicspath{{media/}}     
\usepackage{amsmath}
\usepackage{caption}
\usepackage{subcaption}
\title{Exploring the Electronic and Mechanical Properties of TPDH-Nanotube: Insights from Ab initio and Classical Molecular Dynamics Simulations
}

\author{
  Juan Gomez Quispe, Pedro Alves da Silva Autreto \\
  Center of Natural and Human Sciences, Federal University of ABC \\
  Santo Andre \\
  Sao Paulo, Brazil\\
  \texttt{pedro.autreto@ufabc.edu.br} \\
   \And
   \\ Douglas Soares Galvao \\ 
   \texttt{galvao@ifi.unicamp.br} \\
  Applied Physics Department, State University of Campinas, Campinas, Sao Paulo, Brazil \\
}

\begin{document}
\maketitle

\begin{abstract}
Tetra-Penta-Deca-Hexa-graphene (TPDH) is a new 2D carbon allotrope with attractive electronic and mechanical properties. It is composed of tetragonal, pentagonal, and hexagonal carbon rings. When TPDH-graphene is sliced into quasi-one-dimensional (1D) structures like nanoribbons, it exhibits a range of behaviors, from semi-metallic to semiconducting. An alternative approach to achieving these desirable electronic (electronic confinement and/or non-zero electronic band gap) properties is the creation of nanotubes (TPDH-NT). In the present work, we carried out a comprehensive study of TPDH-NTs combining  Density Functional Theory (DFT) and Classical Reactive Molecular Dynamics (MD). Our results show structural stability and a chiral dependence on mechanical properties. Similarly to standard carbon nanotubes, TPDH-NT can be metallic or semiconductor. MD results show Young's modulus values exceeding $700$ GPa, except for nanotubes with very small radii. However, certain chiral TPDH-NTs (n,m) display values both below and above $700$ GPa, particularly for those with small radii. The analyses of the angle and C-C bond length distributions underscore the significance of the tetragonal and pentagonal rings in determining the mechanical response of TPDH-NTs (n,0) and (0,n), respectively.
\end{abstract}

\keywords{Nanotube \and Molecular Dynamics, Density Functional Theory}

\section{INTRODUCTION}

Graphene stands out as being presently one of the most investigated materials due to its unique electronic and mechanical properties and potential wide applications, including medicine, electronic devices, agriculture, etc. \cite{Khan2019,Olabi2021,Mbayachi2021}. The distinctive feature of graphene lies in its two-dimensional topology, where $sp_{2}$ covalently bond carbon atoms in hexagonal rings form a triangular lattice, resulting in high resistance to mechanical deformations and high electronic mobility \cite{Geim2009,Memarian2015}.\\

Topologically, graphene nanoribbons (GNRs) and carbon nanotubes (CNTs) are structures formed by cutting and rolling up graphene layers. GNRs can be considered as quasi-1D carbon strips with finite-sized $sp2$ hybridization and well-defined (armchair or zigzag) edges  \cite{Taherpour2018a}. Depending on the GNR size, the electronic confinement can result in a non-zero electronic bandgap, while graphene\cite{Dutta2010} has a null bandgap. CNTs depending on their chirality, can exhibit semiconductor or metallic behavior \cite{Dai2021,Rafii-Tabar2007}.\\

Graphene is just one member of the 2D carbon allotrope family. There are hundreds of proposed structures, although only a few of them have been experimentally realized \cite{Jana2022}. 2D tetra-penta-hepta graphene (TPH-gr) is one of such allotropes that has recently been synthesized by the dehydrogenative C-C coupling of 2,6-polyazulene chains \cite{Fan2019c}. TPH-gr is composed of $C_{4}, C_{5}$, and $C_{6}$ carbon rings, having two phases exhibiting semiconductor properties with direct electronic band gaps of 2.704 and 2.361 eV, respectively \cite{Zhang2021c}. A closely related structure is the 2D tetra-penta-deca-hepta graphene (TPDH-gr). TPDH-gr is composed of $C_{4}, C_{6}, C_{10}$ and $C_{6}$ carbon rings (see Figure \ref{fig:tpdh_systems}). TPDH-gr was proposed by D. Bhattacharya and D. Jana \cite{Bhattacharya2021}.\\ 

Density Functional Theory (DFT) calculations indicate that TPDH-gr exhibits a semi-metallic behavior. However, when TPDH-graphene is in the form of nanoribbons, confined in a quasi-1D structure, they present different electronic properties, such as a direct electronic bandgap of $\sim 2.0$ eV for one of the nanoribbon types (NR3) \cite{Bhattacharya2021}. It was reported that TPDH-gr has anisotropic mechanical properties, with values of stiffness constants of $C{11} = 244.48$ Gpa.nm, $C_{22} = 366.51$ Gpa.nm and $C_ {12} = 62.15$ Gpa.nm, indicating that TPDH-graphene is stiffer when deformed along one direction than the other. It was also reported that TPDH-gr could be functionalized with hydrogen atoms, changing its electronic properties and creating anisotropy in its electronic transport \cite{D3CP00186E}. Additionally, through molecular dynamics simulations, C. Oliveira et al. \cite{D3CP00186E} showed that the carbon rings $C_{4}$ are the sites of preferential hydrogenation (up to $80 \%$ at room temperature).\\

The aforementioned studies illustrate the dependence of electronic and mechanical properties on the types of carbon rings that compose these allotropes. This is clearly evident in the cases of TPH-gr and TPDH-gr, which exhibit quite distinct electronic properties. Furthermore, electronic confinement within one-dimensional (1D) structures, such as nanoribbons and nanotubes, also introduces significant changes in both the electronic and mechanical properties. Although nanoribbons of diverse carbon allotropes have been extensively investigated, the current literature still lacks such studies on the investigation of their corresponding nanotubes \cite{Pereira2023,Wang2020b,Brandao2021,Sharma2023,Xu2020a}.\\

In this work, we have investigated the electronic and mechanical properties of TPDH-gr nanotubes (TPDH-NTs) through density functional theory (DFT) and classical reactive molecular dynamics (MD) simulations. The TPDH-NTs considered in this work include zigzag ((n,0), (0,n)), armchair (n,n), and some chiral (n,m) nanotubes. The DFT simulations were used to obtain the electronic band structures and their corresponding density of states, to determine the semiconductor or metallic nature of the nanotubes. MD was used to obtain the stress-strain curves and some elastic properties such as Young's modulus ($Y_{M}$), ultimate strength (US), and fracture limit (FL) values. Furthermore, the deformation mechanisms were investigated to address the relative importance of the different carbon rings on the TPDH-NT mechanical properties.\\

\section{COMPUTATIONAL METHODS}
\subsection{Modeling of TPDH-NTs}

\begin{figure}[h]
    \centering
    \includegraphics[width=0.5\textwidth]{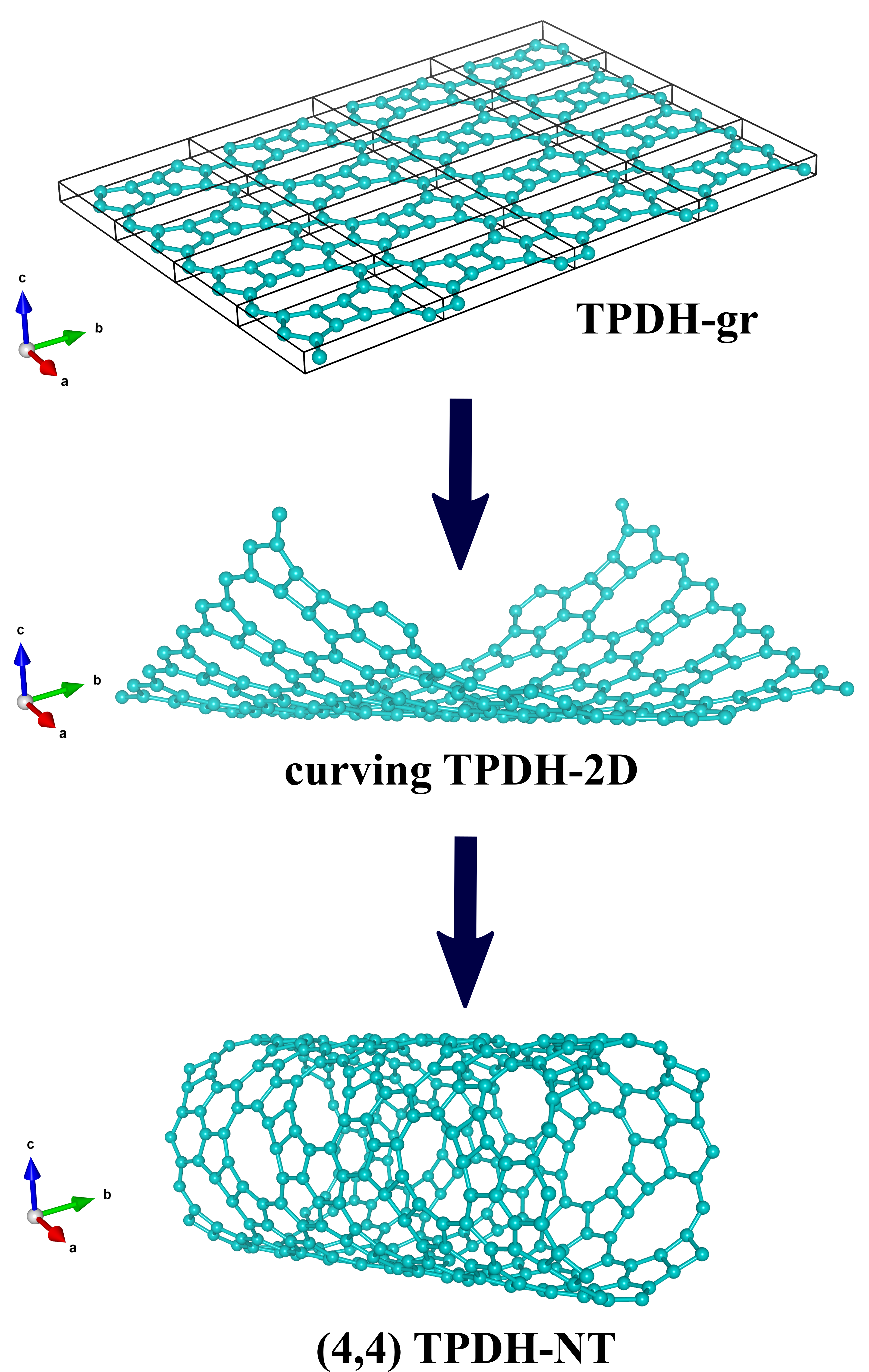}
    \caption{Representative scheme of generating the (4,4) TPDH nanotube. Top: TPDH-gr unit cells replicated for $n = 4$ and $m = 4$, where $n$ and $m$ are the chiral indices of the obtained TPDH nanotubes. Middle: TPDH-gr curving process. Bottom: Perspective view of formed TPDH-(4,4) nanotube. The preservation of the TPDH-gr native carbon rings ($(C_{4},C_{5},C_{6},C_{10})$) can be observed.}
    \label{fig:createtpdh}
\end{figure}

As mentioned above, the TPDH-gr is composed of tetragonal $(C_{4})$, pentagonal $(C_{5})$, hexagonal $(C_{6})$, and decagonal $(C_{10})$ rings, as shown in Figure \ref{fig:createtpdh}. The TPDH-gr primitive vectors of the unit cell $(\vec{p} = m \vec{a} + n \vec{b})$ are also shown for the m, n = 4 case.

To create the TPDH-NTs models, we used the Cif2Tube code \cite{Wang2016}, which uses the crystallographic information (\textit{*.cif} files) for generating the corresponding nanotube or nanoscroll-type systems. Cif2Tube applies the following transformation operations to the atomic positions of TPDH-gr
in order to generate a nanotube-like system \cite{Wang2016}:

\begin{equation}
\begin{split}
    x_{i}^{\prime} &= (R + k \cdot z_{i}) \times \cos(-k \cdot \frac{x_{i}}{R}) \\
    y_{i}^{\prime} &= (R + k \cdot z_{i}) \times \sin(-k \cdot \frac{x_{i}}{R}) \\
    z_{i}^{\prime} &= -y_{i}, \\
\end{split}
\label{eq:cif2tube}
\end{equation}

where $x_{i}$, $y_{i}$, and $z_{i}$ are the TPDH-gr coordinates of the $ith$ atom, $R$ is the radius of the TPDH-NT, $k$ is a dimensionless parameter that can take the values of $1$ or $-1$, and $x_{i} ^{\prime}$, $y_{i}^{\prime}$ and $z_{i}^{\prime}$ are the coordinates of the $ith$ atom in the created nanotube system. Designation of the external surface is achieved using the value of $k = [ ( m
a , n b ) \times \Vec{L} ] \cdot \Vec{c}$, where the vector $\Vec{L}$ is perpendicular to $( m \Vec{a} , n \Vec{b} )$, and $\Vec{c}$ is a vector parallel to $\Vec{z}$. If the value of $k$ is positive, then the top surface becomes the external surface of a nanotube or a nanoscroll and vice versa.\\

\begin{figure}[h]
    \centering
    \includegraphics[width=0.8\textwidth]{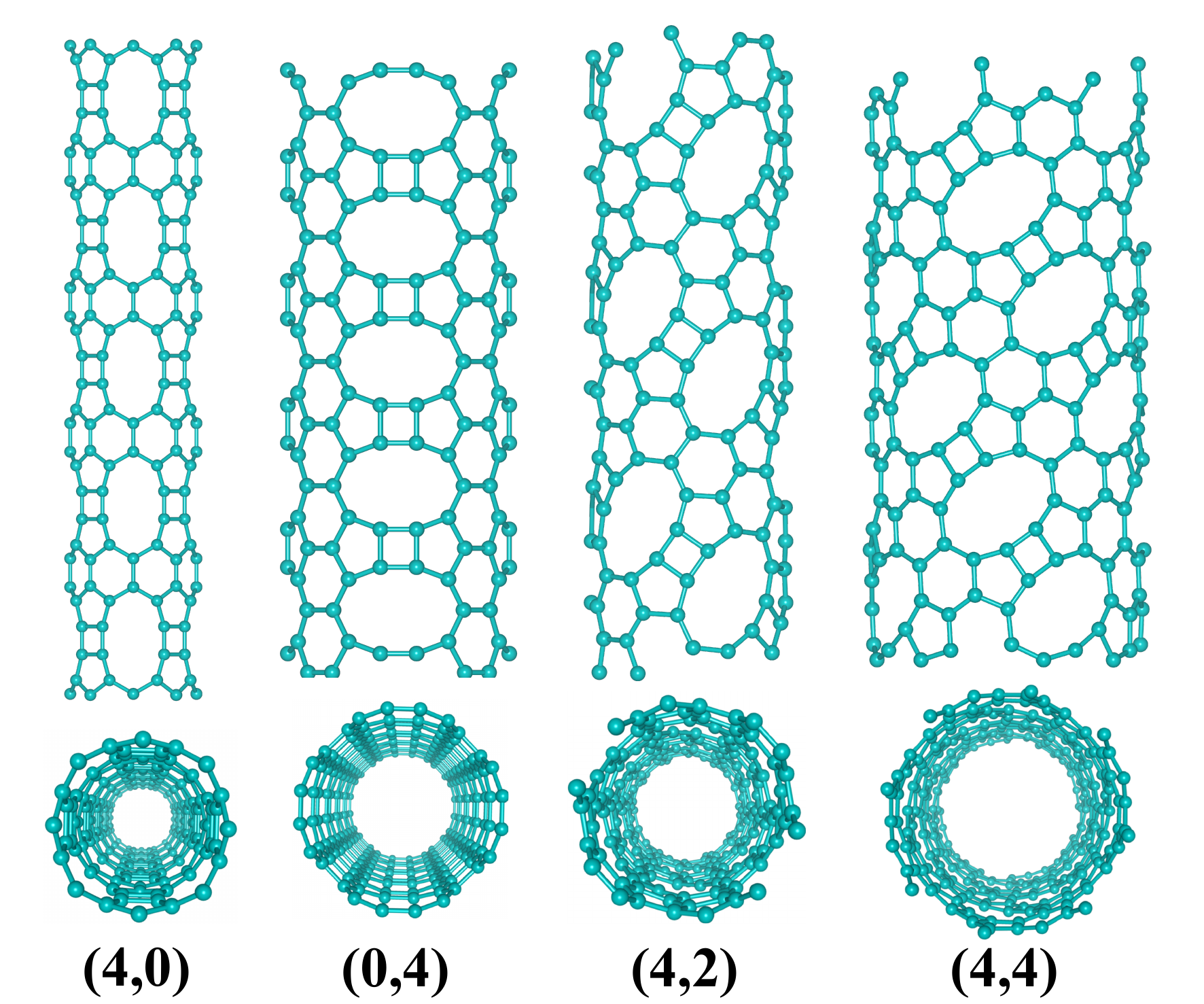}
    \caption{Representative examples (top and perspective views) of the types of TPDH nanotubes considered in this work. Zigzag  ($(4,0)$ and $(0,4)$), chiral ( $(4,2)$) and armchair ($(4,4)$).}
    \label{fig:tpdh_systems}
\end{figure}

We have considered four types of TPDH-NTs: zigzag, inverse zigzag, armchair, and chiral, which can be seen in Figure \ref{fig:tpdh_systems}.

\subsection{Ab initio simulations}
In this work, we carried out DFT \cite{Kohn1965,Hohenberg1964} simulations using the SIESTA code \cite{Soler2002a,Garcia2020} to investigate the structural stability and electronic properties of TPDH-NTs. Kohn-Sham orbitals were expanded using a double-$\zeta$ basis set composed of numerical pseudo-atomic orbitals of finite range enhanced with polarization orbitals. A GGA-PBE type exchange and correlation functional was used, and to define the cutoff radii of the basis function, a common atomic confinement energy displacement of $0.02$ Ry was applied, while the real-space grid fineness was determined by a mesh cutoff of $350$ Ry \cite{Anglada2002}. Regarding the exchange-correlation potential, the generalized gradient approximation was chosen \cite{Perdew1996}, and pseudopotentials were modeled within the conservative Troullier-Martins \cite{Troullier1991} norms in the Kleinman-Bylander factorized form \cite{Bylander1982}. \\

The created structural TPDH-NT models were fully relaxed (residual forces smaller than 0.01 eV/Angstrom), and we adopted a convergence criterion where self-consistency is achieved when the maximum difference between the output and the input of each element of the density matrix is less than $10^{-4}$ eV. The Brillouin zone integrations were carried out using a Monkhorst-Pack \cite{Hu2019} grid of $1 \times 1 \times 8$ k-points. Periodic boundary conditions were imposed with lattice vectors perpendicular to $a_{x}$ and $a_{y}$, and with a sufficiently large periodic boundary ($\sim 20$ \AA) to simulate the vacuum and prevent spurious interactions between periodic images.\\

The cohesive energy $E_{\text{coh}}$ (eV) was calculated for each TPDH-NT using the following equation:
\begin{equation}
    E_{\text{coh}} = (E_{\text{total}} - n E_{\text{gas}})/(n)
\end{equation}
where \(E_{\text{total}}\), \(n\), and \(E_{\text{gas}}\) represent the total energy, number of carbon atoms, and energy of isolated carbon atoms, respectively. A more negative \(E_{\text{coh}}\) corresponds to a more stable structural configuration.\\

When a tubular structure is formed from a TPDH-gr, an internal strain is generated, this strain energy being specified as the Curvature Difference Energy $E_{\text{curv}}$ \cite{Gulseren2002}. This parameter is crucial for assessing the stability of the tube concerning the corresponding infinite flat sheet \cite{Koch2015a,Xu2020a}. The curvature energy, \(E_{\text{curv}}\), is calculated as follows:
\begin{equation}
     E_{\text{curv}} = E_{\text{tube}} - E_{\text{sheet}} 
\end{equation}
where \(E_{\text{tube}}\) and \(E_{\text{sheet}}\) represent the total energy of the TPDH-NT and its corresponding TPDH-gr (sheet), respectively.\\

In the classical theory of elasticity, the curvature energy is expressed using the following formula \cite{Robertson1992,Gulseren2002}:
\begin{equation}
    E_{\text{curv}} = \frac{Y h^{3} \Omega }{24 R^{2}} = \frac{\alpha}{R^{2}} \\
    \label{eq:clasiclimit}
\end{equation}
where $Y$ is the Young's modulus, $\Omega$ is the area per carbon atom, $R$ is the radius of the tube, and $h$ is the thickness of the tube's wall. Therefore, the approximate calculation of $\alpha$ can be quite useful to compare Young's modulus of different TPDH-NTs.

\subsection{Molecular dynamics simulations}

To investigate the mechanical properties of TPDH-NTs, we carried out classical reactive MD simulations at room temperature ($T = 300K$) using a reactive force field (ReaxFF) implemented in the LAMMPS code \cite{Kryuchkov2018}. \\

The selected TPDH-NTs have an initial length (along the $z$ direction) of approximately $100$ \AA\ for all types of TPDH-NTs (zigzag, inverse zigzag, armchair, chiral) considered here.
Considering the number and size of the TPDH-NTs, to carry out this study with ab initio DFT molecular dynamics would be computationally very expensive.\\

To remove any initial stress before the stretching procedure, we subjected the nanotubes to a thermalization protocol within an isothermal-isobaric ensemble \cite{Evans1983}, with the pressure set to zero. Throughout all MD simulations, the temperature was maintained at a constant $300$K, controlled by a Nosé-Hoover thermostat \cite{Hoover1985}.\\

The stretching is induced by increasing the simulation box size along the periodic direction (the $z$-axis). The system dynamics were updated for every increase in $0.25$ fs, maintaining a constant deformation/elongation rate of $\sim 3.5 \times 10^{-6}$ fs$^{-1}$. The elastic properties were evaluated using Young's modulus values, estimated as ($Y = d\sigma_{ii}/d\epsilon_{ii})$, where $(\sigma_{ii})$ represents the component of the virial tensor stress, and ($\epsilon_{ii}$) denotes the deformation along the axial direction $i$. The stress tensor is defined as:
\begin{equation}
    \sigma_{ij} = \frac{\sum_{k}^{N} m_{k} \nu_{k_{i}} \nu_{kj} }{V} + \frac{\sum_{k}^{N} m_{k} r_{k_{i}} \cdot f_{kj} }{V},
\end{equation}

where $V = Ah = L_{0} \pi d_{t} h$, considering a hollow cylinder, is the volume of the TPDH-NT (zigzag, armchair, chiral), $L_{0}$ and $d_{t}$ are the length and the diameter of the TPDH-NT, respectively, and $h = 3.35$ \AA \  is the thickness value of TPDH-NT structure. The stress distribution per atom in space was calculated using the von Mises stress tensor ($\sigma_{VM}$), which is defined as \cite{Hao2021,DeSousa2019,Rafii-Tabar2007} :
\begin{equation}
    \sigma_{VM}^{i} = \sqrt{\frac{ (\sigma_{xx}^{i} - \sigma_{yy}^{i})^{2} + (\sigma_{yy}^{i} - \sigma_{zz}^{i})^{2} + (\sigma_{xx}^{i} - \sigma_{zz}^{i})^{2} + 6 [ (\sigma_{xy}^{i})^{2} + (\sigma_{yz}^{i})^{2} + (\sigma_{zy}^{i})^{2} ]}{2}}.
    \label{eq:von_mises}
\end{equation}

The average spatial distribution of the atomic stress evolution for all TPDH-NTs during the stretching process is also determined using the von Mises stress tensor (Eq. \ref{eq:von_mises}). This stress tensor is specifically used to qualitatively evaluate the accumulation and/or dissipation of the stress values within the stretched structures.

\section{RESULTS AND DISCUSSIONS}
\subsection{Structural Stability and Electronic Properties}
The arrangement of atoms in a TPDH-gr sheet is shown in Figure S1 in the Supplementary Material. The lattice constants for the primitive unit cell are defined as $\Vec{a}_{1} = 4.97$ \AA\ and $\Vec{a}_{2} = 7.02$ \AA. This primitive unit cell contains $12$ carbon atoms, categorized into four non-equivalent types (denoted as $1$, $2$, $3$, and $4$ in Figure S1). Consequently, seven types of bonds are formed, including $C_{1}^{s}-C_{1}^{s}$, $C_{1}^{L}-C_{1}^{L}$, $C_{1}-C_{2}$, $C_{2}-C_{3}$, $C_{2}-C_{4}$, $C_{3}-C_{3}$, and $C_{4}-C_{4}$ bonds. The $C_{1}^{s}–C_{1}^{s}$ bonds denote connections between carbon atoms of type $C_{1}$ with a bond distance of $1.45$ \AA, while the $C_{1}^{L}-C_{1}^{L}$ bonds involve carbon atoms of type $C_{1}$ with a larger bond length of $1.50$ \AA. The other types of bonds, $C_{1}-C_{2}$, $C_{2}-C_{3}$, $C_{2}-C_{4}$, $C_{3}-C_{3}$, and $C_{4}-C_{4}$, exhibit lengths of 1.42, 1.45, 1.44, 1.37, and 1.48 \AA, respectively. Our calculated lattice constants and bond length values are in good agreement with those reported in the literature \cite{Bhattacharya2021a}.\\

Our calculations showed a cohesion energy value of $E_{\text{coh}} = -9.591$ eV for TPDH-gr, just $0.348$ eV larger than the cohesive energy for graphene. This result indicates that the TPDH-gr has cohesive energy values in the same range of graphene and other 2D carbon allotropes, such as $\alpha$ -graphene, graphdiyne, and twin T-graphene \cite{Jana2022}. \\

With the lattice parameters optimized for TPDH-gr, we created the four structural TPDH-NTs models (zigzag, armchair, and chiral) by rolling up the TPDH-gr sheets. As explained in the computational details section, the rolling was carried out using the Cif2Tube software \cite{Wang2016a}, which uses the mathematical transformations shown in Eq. \ref{eq:cif2tube}, and the TPDH-NTs generated are described by the indices (n,m) that are analogous to the chiral indices of carbon nanotubes.\\

\begin{figure}[h]
    \centering
    \includegraphics[width=\textwidth]{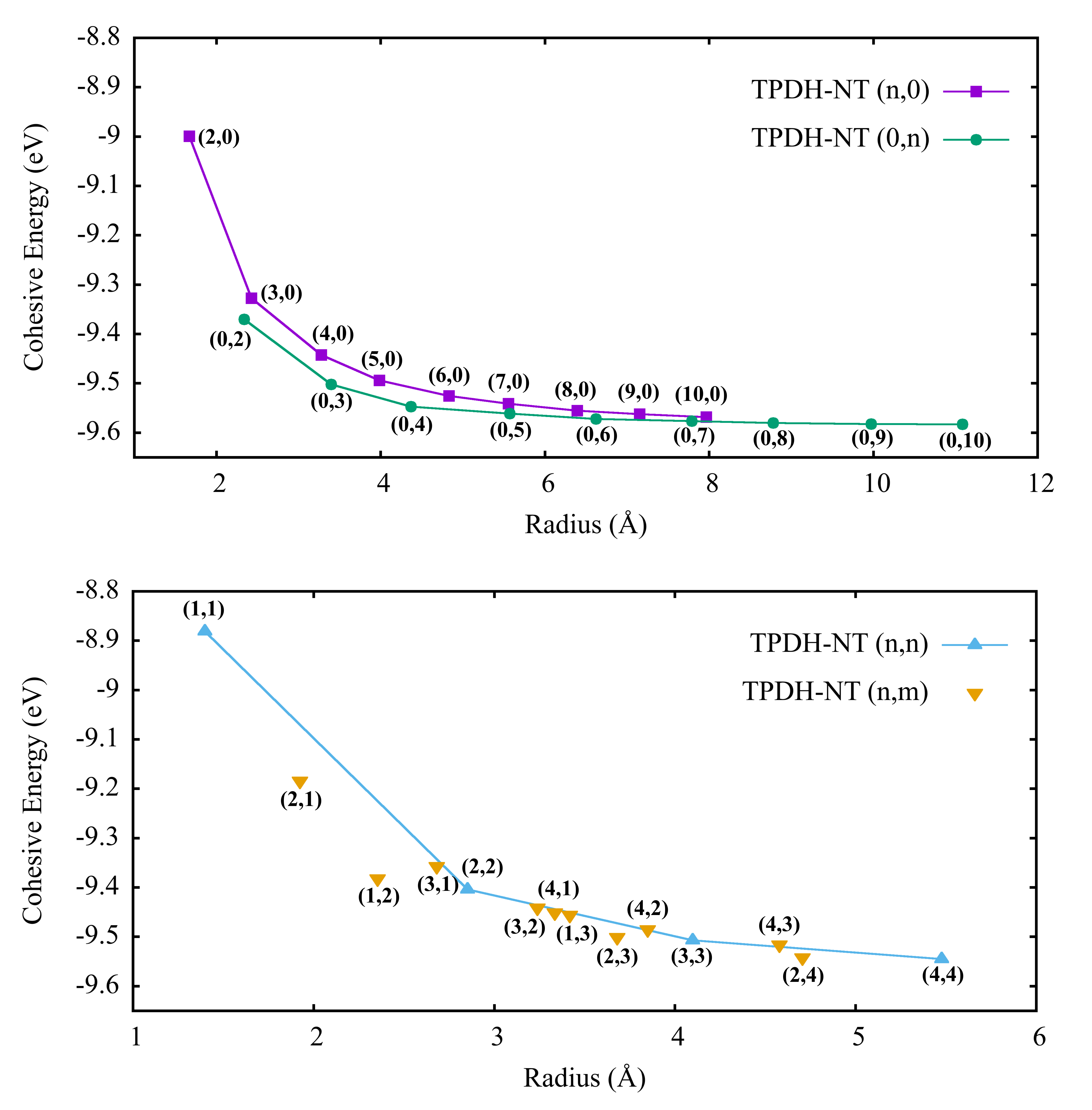}
    \caption{Cohesive energy (eV) values as a function of diameter (Angstroms) for zigzag tubes (Top). Bottom: Armchair (n,n) and Chiral (n,m)TPDH-NTs.}
    \label{fig:cohesive_energy}
\end{figure}

In Figure \ref{fig:cohesive_energy}, we present the values of the cohesive energy as a function of radius for the four TPDH-NT types considered in this work. Due to the high computational cost related to the number of atoms in the minimum unit cell, we have considered zigzag and inverse zigzag TPDH-NTs with a maximum value of the chiral index of $n=10$, armchair TPDH-NTs with a maximum value of the chiral index $n=4$ and some chiral TPDH-NTs.\\ 

As mentioned above, the cohesive energy is a measure of the structural stability of a system; more negative energies indicate more stable structures, and whether the formation of that system is exothermic (thermodynamically favorable) \cite{Majidi2021,Kang2015}. As can be seen from Figure \ref{fig:cohesive_energy}, for the zigzag, armchair, and chiral TPDH-NTs, the cohesive energy exhibits a strong dependence on the radius of the tubes, decreasing until it converges to the value of the cohesive energy of TPDH-gr.\\

The decrease in $E_{\text{coh}}$ is related to the decrease in the surface strain of TPDH-NTs \cite{Kang2015,Majidi2021}. For very large radius values, the energy differences between TPDH-NT and TPDH-gr tend to be zero, as expected. Based on the previous argument, it can be stated that the zigzag TPDH-NTs (0,n) have better structural stability because they have a larger intrinsic radius.\\

\begin{figure}[h]
    \centering
    \includegraphics[width=\textwidth]{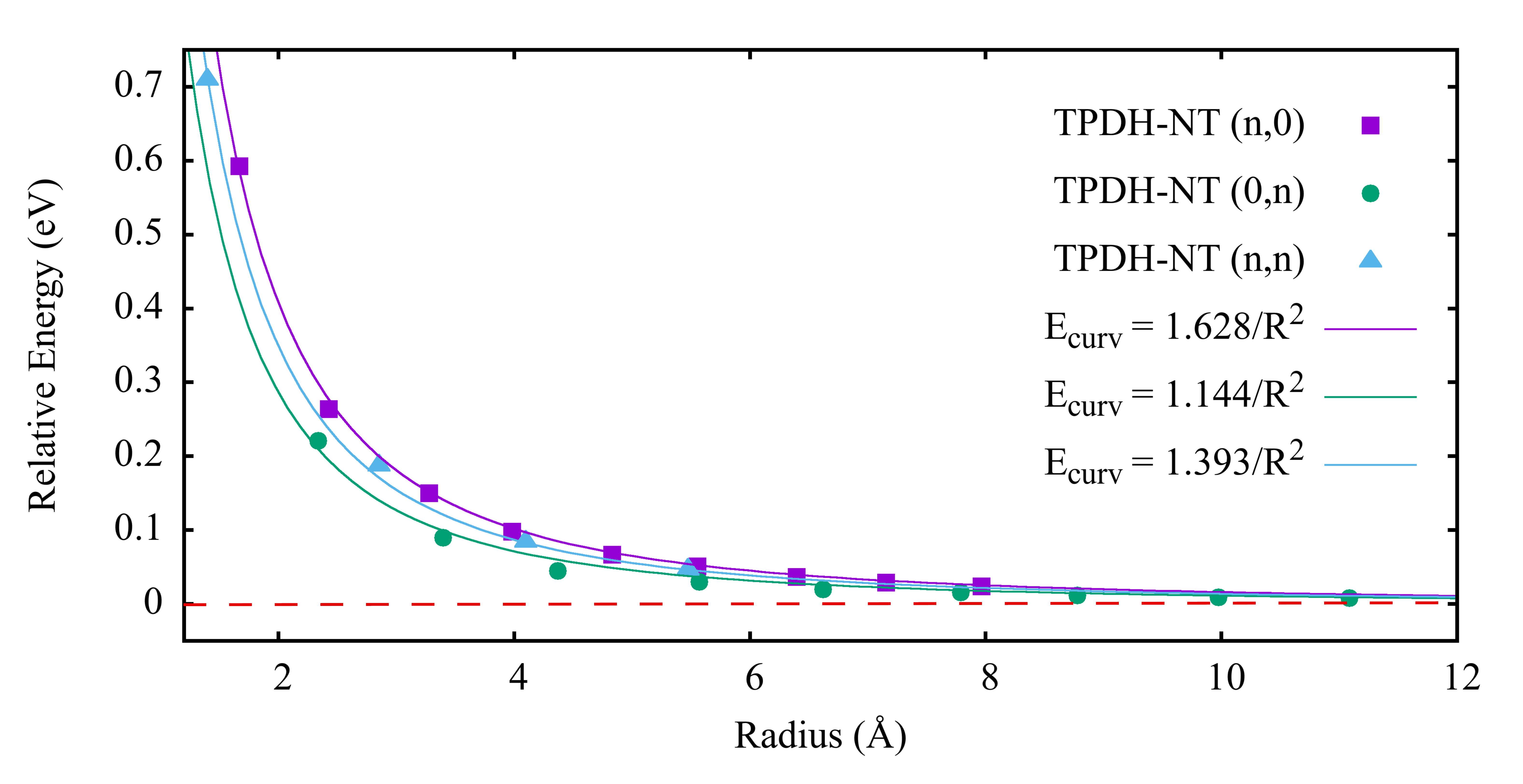}
    \caption{TPDH-CNT relative energy values (eV) for the different tube types. The solid lines represent the fitting of the relative energy using the function: $f(D) = \alpha /r^{2}$.}
    \label{fig:curvature_energy}
\end{figure}

As discussed in the previous section, the relative energy $E_{\text{curv}}$ is a very important parameter because it provides the energy difference between TPDH-NTs and TPDH-gr, and thus be able to determine the effect of chirality and radius on the TPDH-NT structural stability  \cite{Xu2020a,Majidi2021}. In Figure \ref{fig:curvature_energy}, we present the $E_{\text{curv}}$ as a function of the radius for the different TPDH-NTs. We observe that all values are positive values, as expected, since the 
formation of TPDH-NTs is not spontaneous and requires energy to occur \cite{Xu2020a,Majidi2021}. It can also be observed that due to the larger intrinsic radius of the zigzag TPDH-NTs, they require less energy to be rolled up, while the zigzag and armchair TPDH-NTs show almost the same $E_{\text{curv}}$ values.\\ 

In Figure \ref{fig:curvature_energy}, the fitted curves show a dependence on the radius of the NTs of the form: $f(r) = \alpha/r^{2}$. Therefore, we can use Eq. \ref{eq:clasiclimit} to estimate and compare Young's modulus for the TPDH-gr \cite{Gulseren2002,Koch2015}. 

Based on the equation \ref{eq:clasiclimit}, it can be deduced that $\alpha$ is the parameter that contains information about the elastic properties of the sheets that are used to create the nanotubes \cite{Koch2015}. Therefore, we performed a fitting of the $E_{\text{curv}}$ as a function of the radius (R), to estimate the value of the parameter $\alpha$ for the TPDH-NTs. The obtained values are: $\alpha_{(n,0)} = 1.628$, $\alpha_{(0,n)} = 1.144$ and $\alpha_{(n,n)} = 1.393$. These values indicate that the TPDH-gr has a higher Young's modulus along the $\Vec{x}$ direction than $\Vec{y}$. For the TPDH-NTs the longitudinal axis corresponds to $\Vec{x}$ for the zigzag TPDH-NT, while $\Vec{y}$ corresponds to the TPDH-NTs of the inverse zigzag type. Therefore, we can deduce that zigzag TPDH-NTs exhibit a higher resistance to deformation than the other TPDH-NT types. Following this logic, we expect that the armchair TPDH-NTs are the next ordering type of NTs with the highest Young's modulus, while the inverse zigzag type TPDH-NTs are the ones with the lowest resistance to deformation.\\

\begin{figure}[h]
    \centering
    \includegraphics[scale = 1.2]{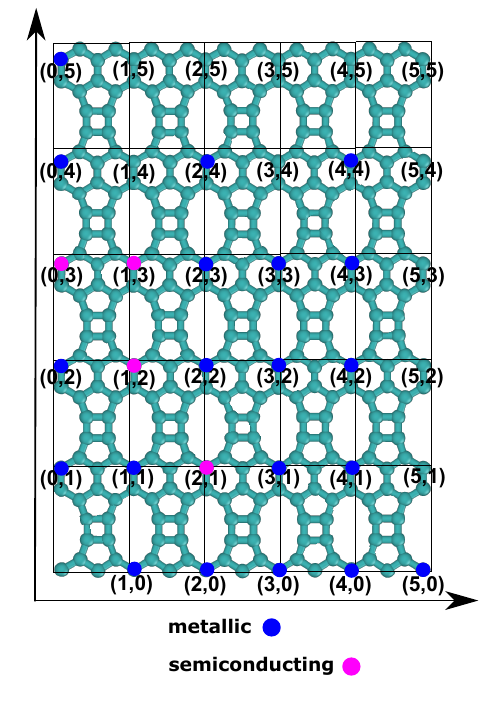}
    \caption{A summary of the electronic behavior (metallic or semiconducting) of the TPDH nanotubes considered in this work.}
    \label{fig:graphmetallicsemiconducting}
\end{figure}

To investigate the electronic properties of the TPDH-NTs, the electronic band structure and density of states of the associated structures were calculated (see Figures S2, S3, and S4 of the Supplementary Materials). TPDH-gr has metallic behavior (no energy gap between the valence and conduction bands), while TPDH-gr nanoribbons exhibit an electronic bandgap indicating semiconductor behavior \cite{Bhattacharya2021}. Due to the 1D confinement effect (nanoribbons) on the electronic properties of TPDH-gr, we can also expect to find a semiconducting behavior for the TPDH-NTs. Figure \ref{fig:graphmetallicsemiconducting} summarizes the metallic or semiconductor behavior of the different TPDH-NTs investigated here. As we can see, the TPDH-NTs with chiral numbers (1,2), (2,1), (0,3), and (1,3) are semiconductors, while the other TPDH-NTs have a metallic behavior. As we can see from Figure \ref{fig:cohesive_energy}, these chiral semiconductor TPDH-NTs are those with small radius values ($< 2.75$ \AA). Another result of interest is the indirect electronic band gap of approximately $0.57$ eV for the TPDH-NT (2,1), which has a radius of $\sim 2.02$ \AA.\\

\subsection{Mechanical Properties}

As mentioned earlier, the study of the mechanical properties of TPDH-NTs was carried out through strain-stress analyses using classical reactive Molecular Dynamics (MD) simulations \cite{Kryuchkov2018}. In Figure S5 are presented the stress-strain curves for TPDH-NTs with zigzag, inverse zigzag, armchair, and chiral tubes ((n,0), (0,n), (n,n), (n,m)), respectively. These results show that TPDH-NTs have distinct mechanical behaviors when under a stretching regime. In tables S1, S2, S3, and S4, some typical characteristics of the stress-strain curves are summarized, such as Young's modulus Gpa ($Y_{M}$), Ultimate Strength Gpa (US), and Fracture Strain $\%$ (FL).\\

Figures S5 (a) and (b) show the stress-strain curves for the TPDH-NTs zigzag (n,0) and inverse zigzag (0,n), as can be seen both types of NTs present different stress-strain profiles, with the values of ultimate strength (US) and fracture limit (FL) being the most distinct characteristic between both NTs, with the TPDH-NTs (0,n) having the highest US values. \\

\begin{figure}[h]
\centering
\begin{subfigure}[b]{0.9\textwidth}
    \includegraphics[width=\textwidth]{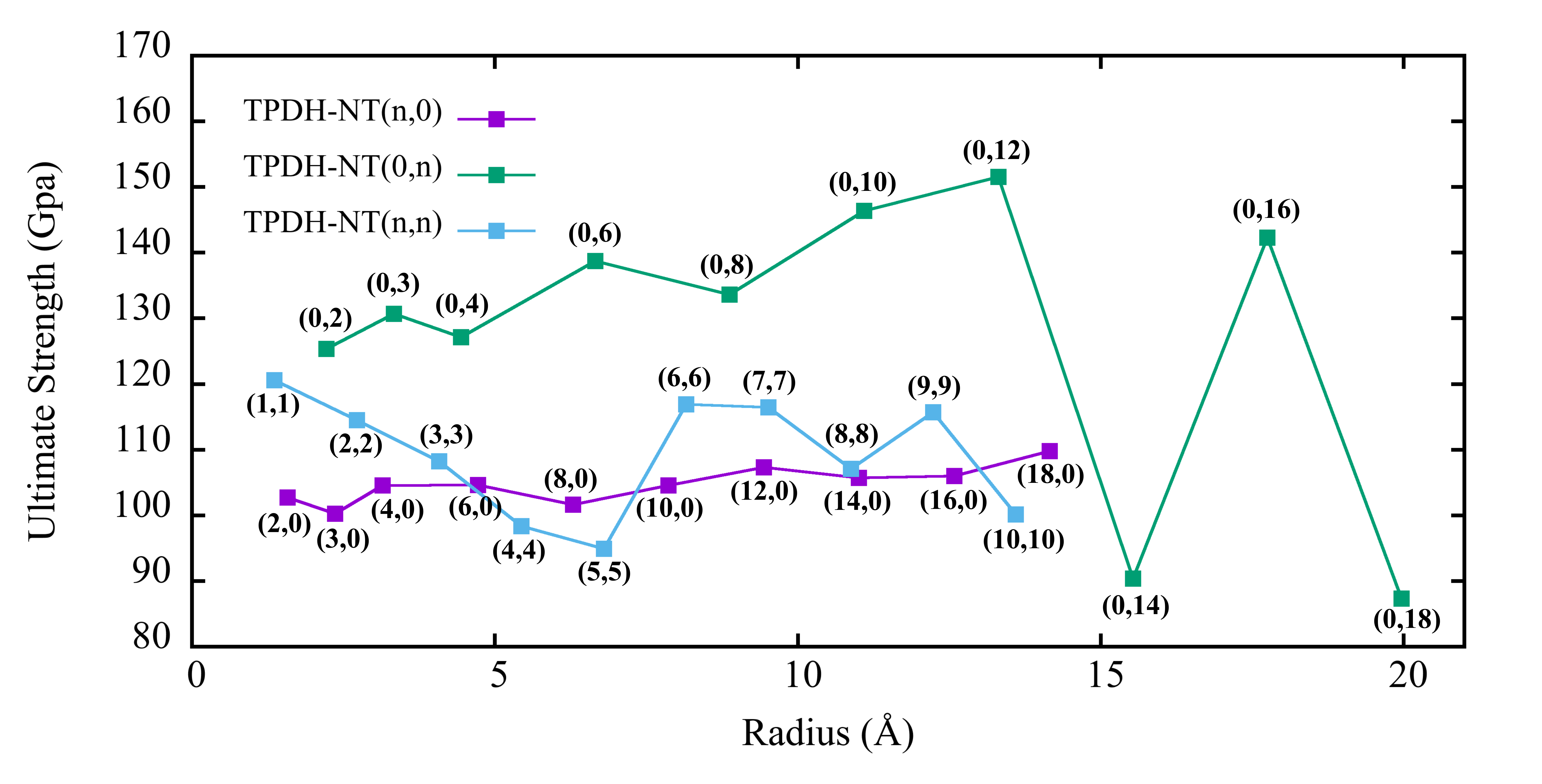}
    \caption{}
    \label{fig:us_tpdh}
\end{subfigure}
\hfill
\begin{subfigure}[b]{0.9\textwidth}
    \includegraphics[width=\textwidth]{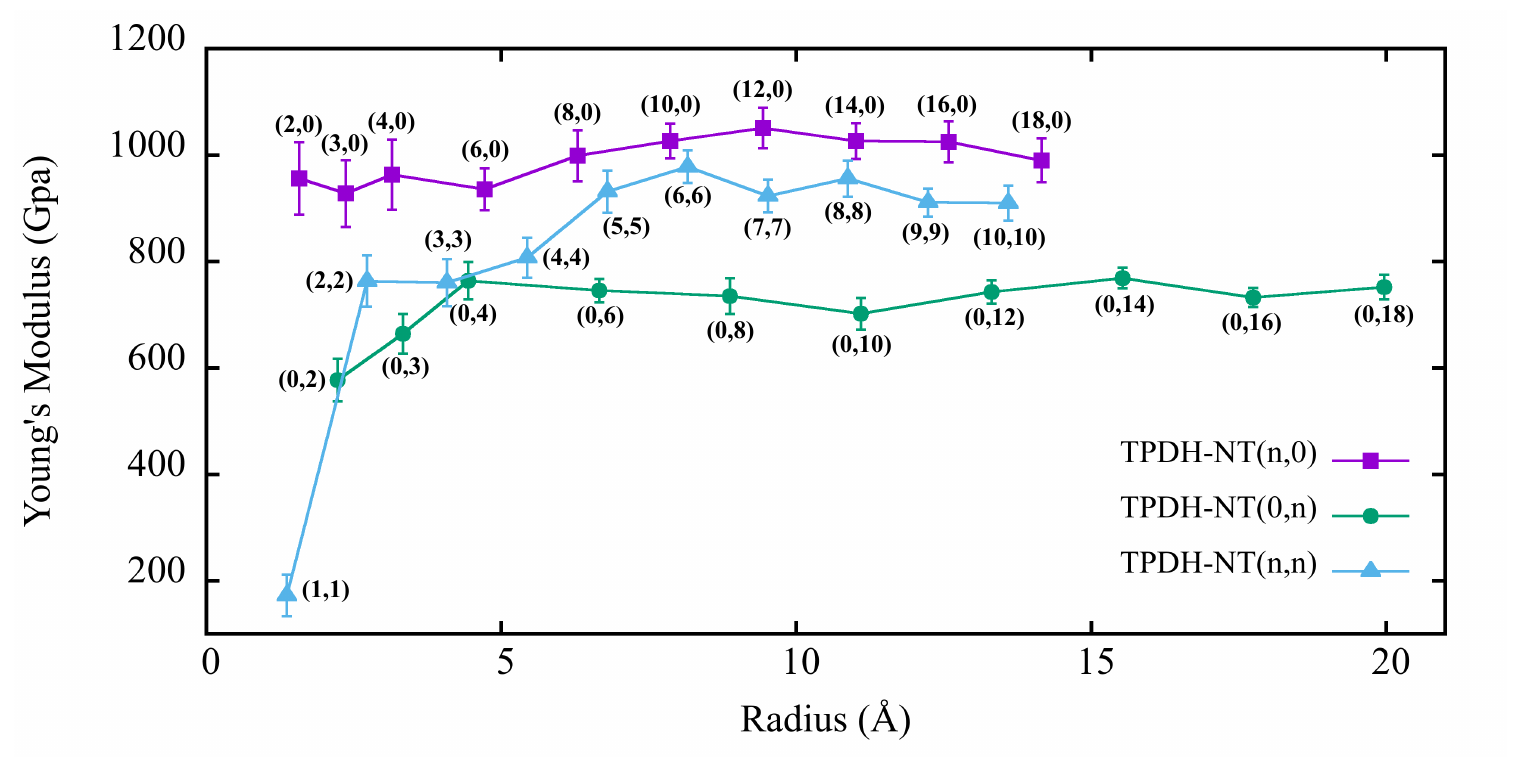}
    \caption{}
    \label{fig:young_module_tpdh1}
\end{subfigure}
\caption{Ultimate Strength (a) and Young's modulus (b) values for TPDH-NTs zigzag (n,0), inverse zigzag (0,n), and armchair (n,n).}
\label{fig:figures}
\end{figure}

From Table S1, we can see that there is a small variation in the ultimate strength (US) values among all zigzag TPDH-NTs (n,0), as can also be seen in Figure \ref{fig:us_tpdh}, where the US values are plotted as a function of the diameter of the nanotubes. Furthermore, a clear correlation is observed between the FL and the nanotube radius. Consequently, nanotubes with larger radii are expected to exhibit FL values exceeding 20.8 $\%$. Conversely, the findings for TPDH-NTs with the chiral index (0,n) (refer to Table S2) indicate high US and FL values that scale with the nanotube radius, reaching a peak at $13.31$ \AA. Beyond this critical radius, larger than $13.31$ \AA, both US and FL tend to decrease and exhibit oscillations (see Figure \ref{fig:us_tpdh}).\\

Table S3 and Figure \ref{fig:us_tpdh} present the FL and US values for the TPDH-NTs with armchair chirality (n,n). It is observed that the US and FL values decrease with the radius compared to the zigzag and inverse zigzag TPDH-NTs. However, for a radius value ($6.80 < r < 12.23$ \AA), then for ($r > 12.23$ \AA), the values of US and FL tend to increase.\\

On the other hand, table S4 shows the values of US, FL, and YM for the chiral TPDH-NTs (n,m). Due to the chiral nature of these NTs, it is not possible to establish a direct relationship of these mechanical parameters with the radius of the NTs. However, the US values shown in Table S4 indicate that TPDH-NTs (n,m) have a wide range of US values with a maximum value for NT (3,1) and a minimum value for NT (2,3).\\

The Young's modulus values for the TPDH-NTs (n,0), (0,n), and (n,n) as a function of the radius of the NT are shown in Figure \ref{fig:young_module_tpdh1}. These results clearly show a dependence of chirality on the elastic properties of the NTs studied. As described by the parameters $\alpha$ coming from the calculations of $E_{\text{curv}}$, the TPDH-NTs with zigzag chirality (n,0) are the stiffest NTs (with the highest value of $Y_{M}$ ). On the other hand, despite being counter-intuitive, TPDH-NTs with inverse zigzag chirality (0,n) are the NTs with the lowest value of $Y_{M}$, indicating that a distribution of horizontal rings $C_{10}$ affects negatively.\\

On the other hand, TPDH-NTs with armchair chirality (n,n) show modulus values of $Y_{M}$ larger than TPDH-NTs (0,n) but lower than TPDH-NTs (n,0), which is in accordance with the classical theory of elasticity. Therefore, we can conclude that the $Y_{M}$ MD results are consistent with the DFT ones (which do not take into account the effect of temperature).\\

\begin{figure}[h]
    \centering
    \includegraphics[scale=0.37]{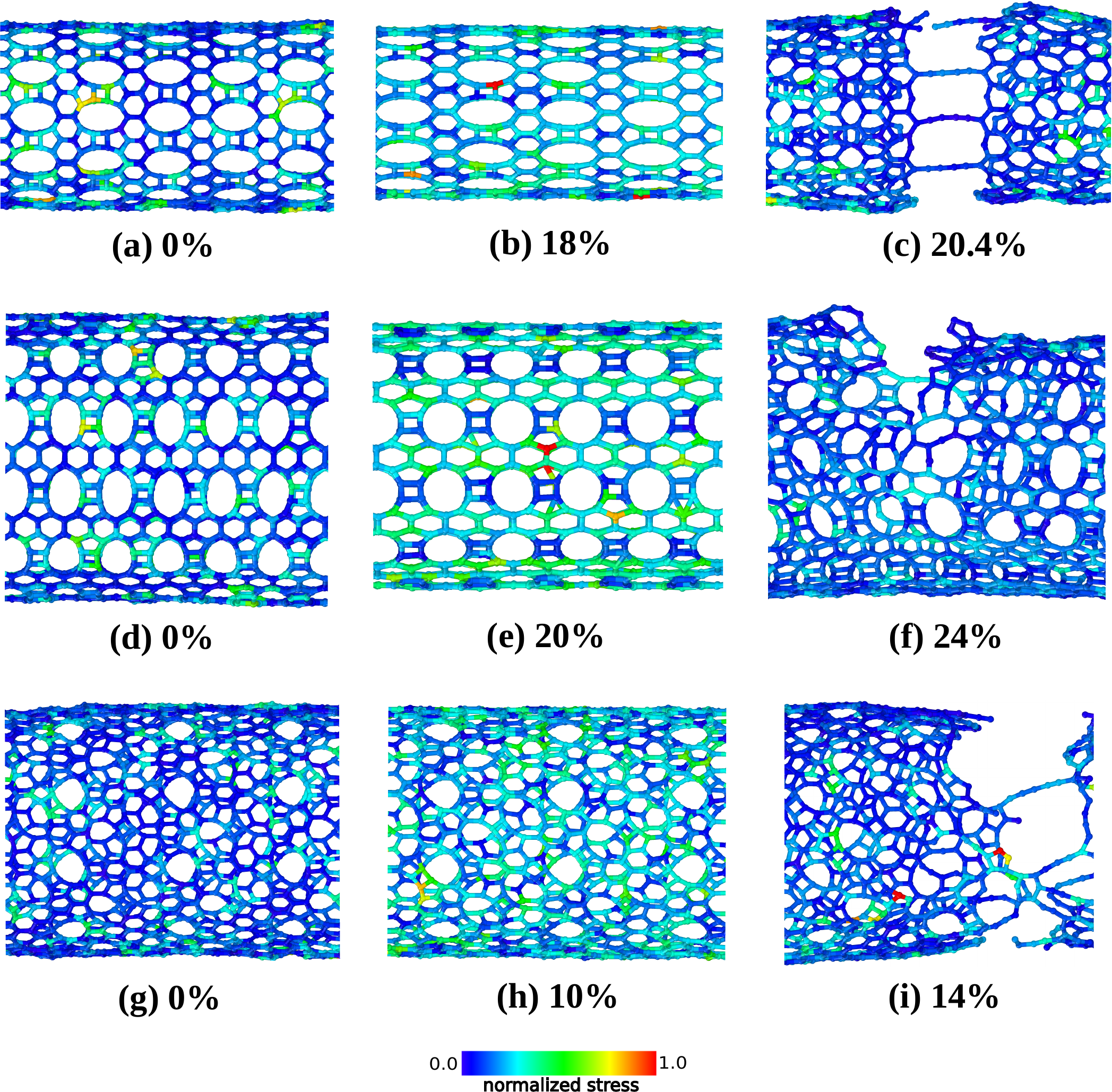}
    \caption{Representative MD snapshots showing the spatial distribution of von Mises stress for the TPDH-NT (zigzag, inverse zigzag, armchair zigzag). (a),(b),(c) TPDH-NT (18,0). (d),(e),(f) TPDH-NT (0,18). (g),(h),(i) TPDH-NT (10,10).}
    \label{fig:von_mises_stess_strain}
\end{figure}

Besides obtaining the values of the mechanical parameters ($Y_{M}$, US, and FL), understanding the TPDH-NT deformation mechanism is also important. Figure \ref{fig:von_mises_stess_strain} shows the spatial distribution of the von Mises stress for the TPDH-NTs (18,0), (0,18), and (10,10) for different strain values. These results help us visualize where the stress accumulates during the deformation process and thus estimate the effect of chirality (structure of the NTs) on these properties.\\

In the top part of Figure \ref{fig:von_mises_stess_strain}, the von Mises distribution of the TPDH-NT (18,0) is shown for the strain of $0\%$, $18\%$ and $20.4\%$. For the value of $0\%$, residual stress is observed from the thermalization process, where apparently the stress is located in the tetragonal $(C_{4})$ and pentagonal $(C_{5})$ type carbon rings.  On the other hand, for a value of $\epsilon = 18\%$, the stress is clearly located in the $C_{4}$ and $C_{5}$ rings. Finally, for $\epsilon = 20.4\%$, the fracture of the NTs is shown, where the breaking of the bonds of the $C_{4}$ rings and the formation of linear carbon chains (LACs) can be observed.\\

Figures S5 and S6 show the distribution and angle values of C-C bonds, respectively, of the TPDH-NT (18,0) for a deformation of $0\%$ and $15\%$. It is possible to observe how the initial angle distribution is affected by strain, shifting the initial angle peak values. The main changes in the angle values involve $C_{6}$ and $C_{5}$, indicating a decrease in the C-bond angle values.\\

\begin{figure}[h]
    \centering
    \includegraphics[scale=0.1]{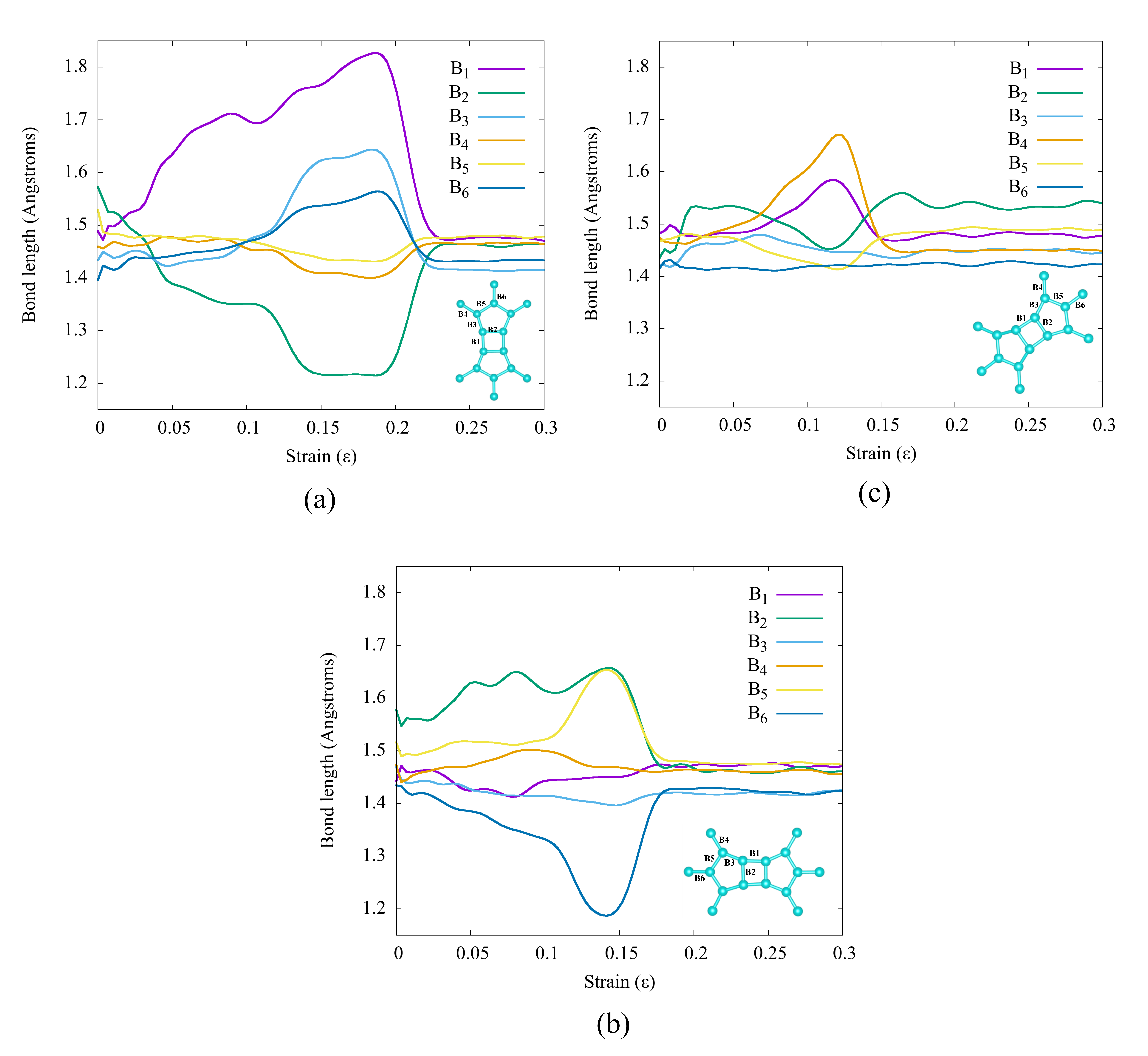}
    \caption{Bond length (C-C) values for TPDH-NT (18,0), TPDH-NT (0,18), and TPDH-NT (10,10). The bonds $B_{1}, B_{2}, B_{3}, B_{4}, B_{5}$, and $B_{6}$ represent equivalent bonds for the TPDH-NTs.}
    \label{fig:bond_lenght_analysis}
\end{figure}

In Figure \ref{fig:bond_lenght_analysis}a, C-C bond distances are presented as a function of strain values. The bonds $B_{1}$ and $B_{2}$ of the tetragonal ring $C_{4}$ exhibit both stretching and compression, occurring in both the elastic and plastic regions. In contrast, bonds $B_{3}$ and $B_{6}$ representing the rings $C_{5}$ and $C_{6}$ only undergo stretching in the plastic region. Consequently, it can be inferred that the tetragonal $C_{4}$ and pentagonal $C_{5}$ rings are the regions where the stress is accumulated. Thus, they play a crucial role in determining the mechanical properties of TPDH-NTs (n,0).\\

Similarly, we can apply the aforementioned analyses to TPDH-NTs with inverse zigzag (0, n) and armchair (n, n) configurations. Figures \ref{fig:von_mises_stess_strain}(d), (e), and (f) present the distribution of von Mises stress for the NT (0,18) at strain levels of $\epsilon= 0\%, 20\%$, and $24\%$. In the case of NT (18,0), residual stress from thermalization is observed. However, at a strain value of $\epsilon = 20\%$, distinctive patterns emerge in the $C_{6}$ ring of NT (0,18), suggesting that stress is accumulated in this region, in contrast to NT (18,0), where it is in the $C_{4}$ rings. This finding is corroborated by the angle distribution analyses, as presented in Figures S5 (b) and S6 (b), showing the angle deformation of the $C_{6}$ ring. Additionally, the analyses of C-C bond lengths indicate stretching and compression in bonds $B_{5}$ and $B_{6}$, which are part of the hexagonal carbon ring $C_{6}$. Therefore, we can conclude that the carbon ring $C_{6}$ present in the NT (0,18) is responsible for this NT having large US and FL values because the carbon bonds in a hexagonal ring are stronger due to their symmetry and hybridization.\\

Finally, the Von Mises stress for TPDH-NTs with armchair chirality (10,10) is presented in Figure \ref{fig:von_mises_stess_strain}(g), (h), and (i) for strain levels of $0\%, 10\%$, and $14\%$. Notably, at a strain of $\epsilon = 10\%$, the stress is more uniformly distributed across the entire nanotube. The C-C angle distribution results (see Figure S5(c)) reveal a more homogeneous profile without prominent peaks. Furthermore, the bond length analyses in Figure \ref{fig:bond_lenght_analysis}c demonstrate the stretching of C-C bonds occurring in almost all carbon rings, with only two unique peaks corresponding to the stretching of $B_{1}$ and $B_{2}$. Based on these findings, we can infer that TPDH-NTs exhibit larger elastic regions compared to other TPDH-NT nanotubes (n,0) and (0,n), which display well-defined yield strength values in their stress-strain curves.\\

\section{CONCLUSIONS}\label{sec4}

In this study, we have investigated the electronic and mechanical properties of TPDH-NTs with chiralities (n,0), (0,n), (n,n), and (n,m) using Density Functional Theory (DFT) and Reactive Molecular Dynamics simulations. The DFT results suggest structural stability, as evaluated by the calculation of the cohesive energy ($E_{\text{coh}}$). The NT structural stability order is $(0,n) > (n,0) > (n,n)$ for radii values less than 5 \AA. This ordering reflects the influence of the surface stress, which decreases as the tube diameter increases. Furthermore, the curvature energy calculations $E_{\text{curv}}$, revealed a chirality dependence in the elastic properties of the TPDH-NT, obtaining values for the alpha parameter, coming from classical elastic theory, in the following decreasing order: $(n,0) > (0,n) > (n,n)$. The chiral TPDH-NTs (n,m) were also investigated, presenting high structural and thermodynamic stability. Simultaneously, we analyzed the electronic band structures and density of states for all types of TPDH-NTs to determine the metallic or semiconductor nature of each nanotube, as summarized in Figure \ref{fig:graphmetallicsemiconducting}.\\

The results from molecular dynamics calculations are consistent with the DFT ones and support the chirality dependence of the elastic properties of TPDH-NTs. Additionally, they indicate that the majority of TPDH-NTs with chiralities (n,0), (0,n), and (n,n) are stiff with $Y_{M}$ values exceeding $700$ GPa, except for NTs with very small radii. However, certain chiral TPDH-NTs (n,m) display $Y_{M}$ values both below and above $700$ GPa, particularly for those with small radii. The analyses of the angle and C-C bond length distributions underscore the significance of the carbon rings $C_{4}$ and $C_{6}$ in influencing the mechanical response of TPDH-NTs (n,0) and (0,n), respectively. In contrast, TPDH-NTs with chirality (n,n) exhibit a more uniform distribution in stress-strain profiles, effectively extending the elastic regime for these nanotubes.\\

The results presented in this study contribute to a deeper understanding of the role played by carbon rings, inherent in the new 2D allotropes and their corresponding nanotubes, on the electronic and mechanical properties of these types of nanostructures.\\

\section{Acknowledgements}

JGQ thanks UFABC Multiuser Computational Center (CCM-UFABC) for computational resources provided. DSG thanks the Center for Computational Engineering \& Sciences (CCES) at Unicamp for financial support through the FAPESP/CEPID Grant 2013/08293-7 and PASA to CNPq (Grant 308428/2022-6).


\bibliographystyle{unsrt} 
\bibliography{references}

\end{document}